\documentclass[aps,prl,twocolumn,showpacs]{revtex4}

\usepackage{amssymb,amsmath,graphicx,graphics,color}

\newcommand{\de}{\partial}

\newcommand{\eq}[2]{\begin{equation} \label{#1} #2 \end{equation}}

\newcommand{\etal}{{\em et al.}}

\newcommand{\pp}{\mathbf{p}}

\newcommand{\AAA}{\mathbf{A}}

\newcommand{\schr}{Schr\"odinger }

\begin{document}

\title{Giant ultrafast Kerr effect in type-II superconductors}
\author{Charles W. Robson}
\author{Kieran A. Fraser}
\author{Fabio Biancalana}
\affiliation{School of Engineering and Physical Sciences, Heriot-Watt University, EH14 4AS Edinburgh, UK}

\begin{abstract}
We study the ultrafast Kerr effect and high-harmonic generation in type-II superconductors by formulating a new model for a time-varying electromagnetic pulse normally incident on a thin-film superconductor. It is found that type-II superconductors exhibit exceptionally large $\chi^{(3)}$ due to the progressive destruction of Cooper pairs, and display high-harmonic generation at low incident intensities, and the highest nonlinear susceptibility of all known materials in the THz regime. Our theory opens up new avenues for accessible analytical and numerical studies of the ultrafast dynamics of superconductors.
\end{abstract}

\maketitle

\paragraph{Introduction ---}

The discovery of superconductivity in 1911 by Kamerlingh Onnes introduced a startling new phenomenon to physics: that of systems exhibiting zero electrical resistance at low temperatures \cite{Annett}. Although superconductivity is relatively easy to measure in the laboratory, a full microscopic explanation was not put forward for over 40 years, in the form of the Bardeen-Cooper-Schrieffer (BCS) theory \cite{BCS}. Between the discovery of superconductivity and its satisfactory explanation there were several powerful attempts: including that of the London brothers in the 1930s \cite{London1}, used to describe the observed expulsion of external magnetic fields; and the Ginzburg-Landau theory of the 1950s \cite{Landau1}, a phenomenological model that was subsequently proved to be strictly linked to the BCS theory in certain limits, and was able to explain the existence of the spatial lattice distribution of flux vortices in type-II superconductors \cite{Abrikosov1}.

Research into superconductivity is currently very active, with recent work on high-temperature superconductors -- reaching up to a critical temperature of $203$ $^{\rm o}$K -- attracting considerable attention \cite{HighT}. The finding that superconductors can act as very effective single-photon detectors also demonstrates their aptness in the field of photonics \cite{Semenov,Gol,Marsili}.

The effect of {\em static} external magnetic fields on a superconductor is one of the subject's most fascinating and studied aspects, and all the introductory and advanced textbooks contain many details on how superconductors behave under the influence of time-independent fields. However, somewhat surprisingly, the influence of a {\em time-dependent} electromagnetic pulse has been somewhat neglected in the literature, with very few exceptions \cite{Eliashberg1,Testardi}. Testardi \cite{Testardi} looked into the experimental destruction of superconductivity by laser pulses; Gor'kov and \'{E}liashberg \cite{Eliashberg1} explored the theory of interaction of weak time-varying fields with impure alloyed superconductors and found tripled-frequency radiation generation -- their derivations utilised Green's functions, complex analysis, and they employed a diagrammatic method. More recently, Matsunaga \etal\ observed experimentally the third-harmonic generation in NbN samples in the THz regime \cite{Matsunaga}, which was explained theoretically by Cea \etal, who calculated the nonlinear currents by using the full microscopic model based on the BCS theory \cite{lara}.

In this paper we present for the first time exhaustive analytical and numerical calculations based on the {\em time-dependent} Ginzburg-Landau equation, and show the appearance of a Kerr effect in a type-II superconductor induced by an incident, arbitrarily short light pulse. We show that, due to progressive and step-like Cooper pair destruction, these superconducting materials display large nonlinear optical behaviour, such as high harmonic generation, at extremely low laser intensities -- typically on the kW/cm$^{2}$  scale in the THz regime  -- and may prove to be a key element in future photonics applications. Our approach is physically more transparent and much more intuitive than the approach based on the BCS Hamiltonian. We specialise our discussion to elemental niobium (Nb) thin films.

\begin{figure}[h]
\centering
\includegraphics[width=8cm]{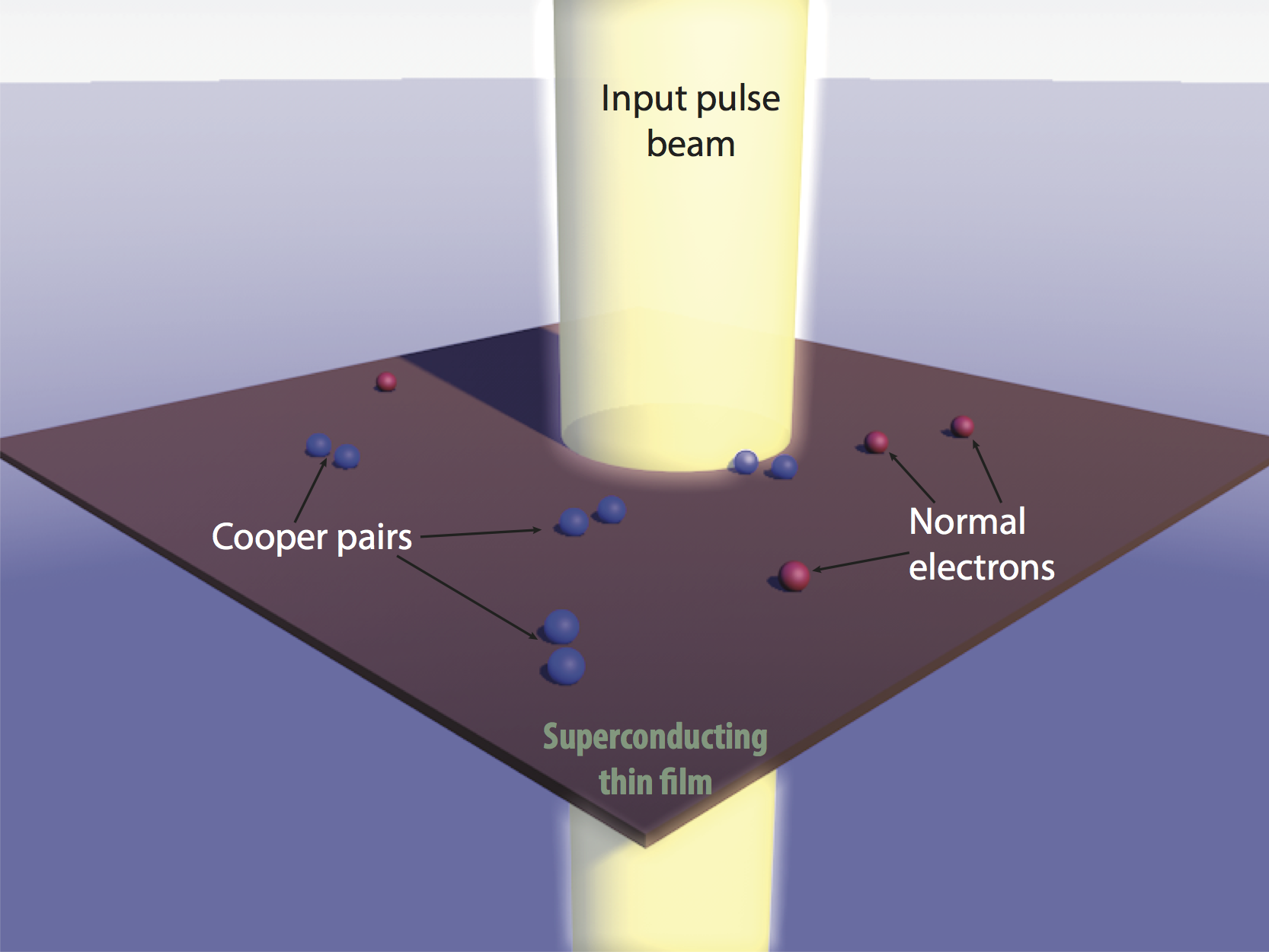}
\caption{Sketch of the system studied in this paper. A superconducting thin film is excited by a normally incident, $x$-polarized electromagnetic pulse. The pulse destroys the Cooper pairs already present in the film, generating more `normal' electrons, which then diffuse and slowly recombine due to the electron-phonon interaction.}
\label{fig1}
\end{figure}

\paragraph{Governing equations ---}

We begin our study with the {\em time-dependent} Ginzburg-Landau equation (TDGLE) \cite{Tang} (using the zero scalar potential gauge):
\begin{equation} \label{eq:TDGL}
\frac{\hbar^2}{2m^*D}\partial_{t}\psi + \frac{1}{2m^*}\left( \mathbf{p}-\frac{q}{c}\mathbf{A} \right)^2 \psi + \alpha\psi + \beta|\psi|^2\psi=0,
\end{equation}
where $\psi(x,y,t)$ is the complex order parameter ($|\psi|^{2}$ is proportional to the total number of Cooper pairs present in the sample: $\psi = \sqrt{n_{\rm c}}e^{i\theta}$), $n_{\rm c}$ is the number density of Cooper pairs, $x,y$ are the coordinates on the plane of the thin film, $m^*$ is the mass of a Cooper pair (approximately equal to twice the mass of a free electron, when neglecting the binding energy), $q=-2e$ is its charge (equal to twice the free electron charge $-e$), $D$ is the diffusion constant, $\AAA(x,y,t)$ is the vector potential describing the spatial and temporal structure of the incident electromagnetic pulse, $\pp$ is the electron momentum operator and $c$ is the speed of light in vacuum. The phenomenological Ginzburg-Landau parameters $\alpha$ and $\beta$ have units of energy and energy$\times$volume respectively and are explicitly given by $\alpha=\alpha_{0}\left(T-T_{c}\right)$, where $\alpha_{0}$ and $\beta$ are constants specific for the material used. $T$ is the temperature of the system and $T_{c}$ is the critical temperature below which superconductivity emerges \cite{Annett}.

Note the quite unconventional structure of Eq. (\ref{eq:TDGL}): since $\psi$ is strictly speaking not a wave function, there is a missing imaginary unit in the first term of this equation, which sets a strong difference from the nonlinear \schr equation. This difference is due to the thermodynamical interpretation of $\psi$ as an order parameter, related to the free energy of the sample \cite{Landau1}. This makes Eq. (\ref{eq:TDGL}) a nonlinear diffusion equation. The time-dependent Eq. (\ref{eq:TDGL}), which is typically not discussed in basic textbooks on superconductivity (where its static version is usually treated), is the core equation of this paper as it describes how the order parameter of a superconductor varies due to an electromagnetic field interaction and contains all of the important system parameters. The Ginzburg-Landau (GL) theory describes the emergence of superconductivity in terms of a phase transition and was shown to be derivable from the BCS theory in close proximity to the critical temperature \cite{Gorkov1}, although the theory is used regularly and has been shown to be practically valid for temperatures well below $T_{\rm c}$. It is a very general phenomenological model used in many areas of physics and avoids the use of many-body wavefunctions \cite{Annett,Grosso}. We choose to use the GL formalism instead of the microscopic BCS theory as we believe the former contains all of the necessary concepts and tools required -- once combined with basic electrodynamics -- to show the ultrafast optical nonlinearities inherent in type-II superconductors. The original motivation for the development of GL theory was indeed to describe superconductivity without needing to delve into its microscopic structure. The gauge field can however be introduced in the BCS theory via a Peierls substitution in real space, as in Ref. \cite{lara}.

The set-up considered in this paper is sketched in Fig. \ref{fig1}. An electromagnetic pulse is normally incident to the 2D superconducting thin film. We make the simplifying assumption that the electric field of the incident light is polarized along the $x$-direction only, as is the magnetic vector potential $\mathbf{A}=[A(x,y,t),0,0]$. Expanding equation (\ref{eq:TDGL}) gives:
\begin{widetext}
\begin{equation} \label{eq:general}
\frac{\hbar^2}{2m^* D}\partial_t\psi-\frac{\hbar^2}{2m^*}{\nabla^2}\psi + \frac{i\hbar q}{2m^*c}(\nabla \cdot \mathbf{A})\psi + \frac{i\hbar qA}{m^*c}(\nabla \psi)+\frac{q^2 A^2}{2m^*c^2}\psi + \beta|\psi|^2 \psi + \alpha\psi=0.
\end{equation}
\end{widetext}

As a first attempt to understand the origin of the optical nonlinearity [not to be confused with the {\em electronic} nonlinearity, which is given by the $|\psi|^{2}\psi$ term in Eq. (\ref{eq:general}), and due to the phonon-mediated electron-electron interaction], we now use Eq. (\ref{eq:general}) to make a rough estimate of the third-order nonlinear optical coefficient exhibited by the thin film. In order to find an intensity scale around which nonlinear optical effects in the superconductor become important we make the following assumptions: the order parameter $\psi$ varies slowly in the spatial variables, and we use the London gauge $\nabla \cdot \mathbf{A}=0$, due to the $\phi=0$ gauge and the fact that at ultrafast time scales the charge density can be taken as zero throughout the process. See the review by Hirsch \cite{Hirsch} for a detailed look into the subtleties associated with the gauge choices and charge density in a superconducting system. Using the expanded TDGLE (\ref{eq:general}), and the assumptions detailed above, gives $\frac{\hbar^2}{2m^* D}\partial_{t}\psi+\left( \frac{q^2 A^2}{2m^*c^2}+\alpha\right) \psi + \beta|\psi|^2 \psi=0$. We see that optical nonlinearities become important when the vector potential term in parentheses is dominant, and by using $\mathbf{E}=-(1/c)\de_{t}\mathbf{A}$ (in the $\phi=0$ gauge) to approximate the amplitudes $A\approx -Ec/\omega_{0}$, along with the usual light intensity equation $I=\epsilon_{0}cE^{2}/2$ (assuming unity refractive index), we find a value for the nonlinear intensity scale given by $I_{0}=c\epsilon_{0}m^{*}|\alpha|\omega_{0}^2/q^2$, where $\omega_{0}$ is the carrier frequency of the pulse. For light of intensity $I\geq I_{0}$ then the term proportional to $A^{2}\psi$ in (\ref{eq:general}), which is responsible for the nonlinear interplay between the field and the superconductor, cannot be ignored; and as we will see below a very powerful cubic nonlinearity is predicted.

The first term in (\ref{eq:general}) represents the rate of change of the order parameter with respect to time; the second term encodes the spatial diffusion dynamics of the Cooper pairs; the third term vanishes due to our choice of gauge for $\mathbf{A}$; the fourth term does not affect the system's nonlinear dynamics, only adding a linear phase; the fifth and sixth terms illustrate the inherent nonlinearities of the system manifested in the electronic cubic order parameter $|\psi|^2\psi$, and the photonic $A^2 \psi$ cross terms; the last term contributes to the ``recovery" of the order parameter's original value, i.e. Cooper pair recombination after the superconductor has interacted with light.

The basic physics of the process giving rise to Kerr optical nonlinearity can be explained as follows: as the light pulse impacts the superconductor Cooper pairs are destroyed, increasing the number of normal electrons in the material. As each progressive peak and trough of the light pulse reaches the material the value of the order parameter drops and then begins to recover as Cooper pairs are destroyed -- generating harmonics -- and then reformed, due to the phonon-mediated potential. This mechanism is reminiscent of the ionization dynamics in plasma physics in which light strips electrons (free electrons are ``created") from their atoms, see for instance a recent work on the subject \cite{fabioplasma}.

The total current in the superconductor is composed of an Ohmic component and a supercurrent, the latter given explicitly by \cite{Tang}:
\begin{equation} \label{eq:supercurrent}
J_{s} = -\frac{q^2}{m^* c}|\psi|^2 A - \frac{iq\hbar}{2m^{*}} \left( \psi^{*}\nabla \psi - \psi\nabla\psi^{*} \right),
\end{equation}
where the second term is proportional to the spatial gradient of the phase $\theta$ of the order parameter, as can be seen if we rewrite (\ref{eq:supercurrent}) as $J_{s} = -\frac{q^2}{m^* c}|\psi|^2 A + \frac{q\hbar}{m^*}|\psi|^2 \nabla \theta$. All time-dependences are implicit and vector notation has been removed as $J_{s}$ has only an $x$-component, since the input pulse is linearly polarized.

\paragraph{Estimate of $\chi^{(3)}$ ---}
The nonlinear electric susceptibility $\chi^{(3)}$ of the superconducting film can be derived as follows. The magnetic vector potential evolves due to the current density via the well-known wave equation:
\begin{equation} \label{eq:box}
\left(\frac{1}{c^2}\frac{\partial^{2}}{\partial t^2}-\nabla^{2}\right)\mathbf{A}=\frac{1}{\epsilon_{0}c}\mathbf{J},
\end{equation}
where $\epsilon_{0}$ is the vacuum permittivity.
Taking the partial time derivative of both sides of equation (\ref{eq:box}) and using $E=-\frac{1}{c}\frac{\partial A}{\partial t}$ produces $\left(\frac{1}{c^2}\frac{\partial^{2}}{\partial t^2}-\nabla^{2}\right)E=-\frac{1}{\epsilon_{0}c^2}\partial_{t}J$, where the right-hand side is proportional to the material polarization. Assuming that the normal part of the current can be disregarded as it does not contribute to the optical nonlinearity (the metallic part of the system is simply absorptive) then the nonlinear polarization is seen to be of Kerr type, i.e. $P_{\rm NL}=\chi^{(3)}E^3$, by solving analytically Eq. (\ref{eq:general}) for a space-independent continuous wave $A(t)=A_{0}\cos(\omega_{0}t)$ and expanding in Taylor series of $A_{0}^{2}$, stopping at the first order. One obtains:
\eq{expansion}{|\psi|^{2}\simeq-\frac{a}{b}-\frac{d}{2b\omega_{0}^3}\left[ a^2 \omega_0 \mathrm{cos}(2\omega_0 t) - a \omega_{0}^2 \mathrm{sin}(2\omega_0 t)  \right] + ...,}
where we have defined $a\equiv2m^* D\alpha/\hbar^{2}$; $b\equiv2m^* D\beta/\hbar^{2}$; $d\equiv q^2 D A_{0}^2/(\hbar c)^{2}$.
Inserting expression (\ref{expansion}) into the supercurrent (\ref{eq:supercurrent}), one can derive the following GL form of the complex nonlinear coefficient:
\begin{equation}
 \chi^{(3)}_{\rm GL}=-\frac{128k_{B}^2 \left( T-T_{c} \right)^2 e^4}{\epsilon_{0}\beta\omega_{0}^6 \pi^2 \hbar^2 m_{e}^2}
\label{chi3}
\end{equation}
The value of $\chi^{(3)}_{\rm GL}$ in Eq. (\ref{chi3}) is negative and depends on the input pulse frequency $\omega_{0}$, and the parameters $\beta$ and $T_{\rm c}$ which depend strongly on the specific material chosen, and of course on the temperature $T$, which must be below $T_{\rm c}$. It must also be noted that the frequency dependence $\omega_{0}^{-6}$ makes the superconductor very nonlinear for longer wavelengths, as it happens, for instance, in 2D Dirac materials like graphene \cite{graphene}. Equation (\ref{chi3}) is the first novel result of this paper.

\paragraph{Simulations and results ---}

Using the above TDGL equation (\ref{eq:TDGL})  we perform time-dependent simulations in both one and two spatial dimensions to analyse how the order parameter of the superconductor evolves under the influence of an ultrashort pulse, which harmonics are produced, and the form of the supercurrent as a function of time. We analyze both cases as although the 1D case is the simplest, and produces the essential results of the paper, we find that the more realistic 2D results illustrate the spatial variation in order parameter with more clarity -- we also expect future experimental tests of the theory to be carried out in a thin-film (i.e. quasi 2D) geometry. The time-dependent simulations are also necessary to check the ultrafast response of the thin superconductive film in realistic situations.

\begin{figure}[h]
\centering
\includegraphics[width=8cm]{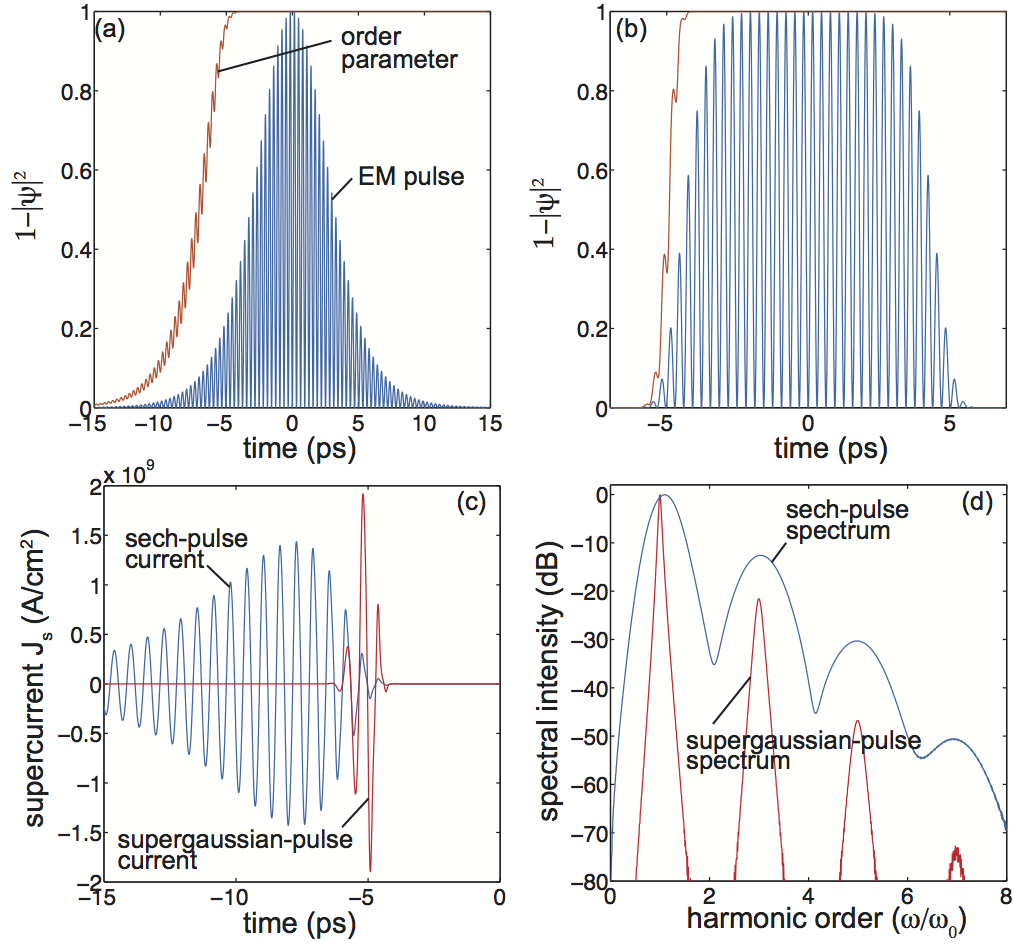}
\caption{(a) Effect of a sech-type pulse on order parameter, showing progressive step-like destruction of Cooper pairs; (b) Same is in (a) but for a supergaussian pulse with same duration and intensity. (c) Supercurrents for the two types of pulses used; (d) Final spectra showing the generation of odd harmonics. The pulse temporal profiles are normalized to unity. Note that in (a,b) $1-|\psi|^{2}$ is plotted, so when Cooper pairs are destroyed this quantity increases towards unity.}
\label{fig2}
\end{figure}

As a representative example, for our estimates we choose pure niobium (Nb) as it has the highest critical temperature $T_{c}$ of all elemental superconductors \cite{Piel}. The following results were taken for simulations of the system at temperature $T=4$ $^{\rm o}$K.

The critical temperature of bulk Nb is $T_{c}=9.25$ $^{\rm o}$K \cite{Stromberg}, and the zero-temperature coherence length $\xi (0)=0.74\sqrt{\chi}\xi_{0}$, where $\xi_{0}$ is the intrinsic BCS coherence length and $\chi$ is the Gor'kov parameter \cite{Gennes}. This parameter depends on whether the superconductor is in the ``dirty limit" or the ``clean limit": a mean free electron path satisfying $l\gg \xi_{0}$ (i.e. the clean limit) gives $\chi\approx 1$; the other limit $l\ll \xi_{0}$ gives $\chi\approx 1.33(l/\xi_{0})$. In the following we will assume our system is in the dirty limit as Nb has a mean free path $l=9.5$ nm and $\xi_{0}=38$ nm (Fermi velocity $v_{\rm F}=1.37\times 10^6\mathrm{m}\mathrm{s}^{-1}$ and mean free electron flight time $\tau_{\rm free}=7$ fs) \cite{Mermin,Maxfield}. In this dirty limit $\xi(0)=16.2$ nm, and the diffusion parameter value is $D=8.09\times 10^{-4} \rm{m^2 s^{-1}}$, while the constants $\alpha_{0}\simeq 1.2618\times 10^{-24}$ J/$^{\rm o}$K, and $\beta\simeq 1.256\times 10^{-51}$ J$\cdot$m$^{3}$.
We study the effects on the order parameter for sech-type pulses and for supergaussian pulses on the scale of $t_{\rm FWHM}=9.55$ ps, which is available experimentally in the THz regime \cite{Yeh}.

Using a laser pulse with carrier wavelength $\lambda_{0}=188$ $\mu$m and a Nb film for a target, we find that optical nonlinear effects should emerge at very low incident intensities of the order of kW/cm$^{2}$, and on ultrafast time scales of the order of picoseconds. Specifically, at a temperature $T=4$ $^{\rm o}$K,  $I_{0}\approx 3.1$ kW/cm$^{2}$. Our parameters used give the value of $\chi^{(3)}\approx -4.24\times 10^{-7}$ m$^{2}$/V$^{2}$, which is to the best of our knowledge the highest theoretical value ever predicted in a $\chi^{(3)}$-material in the THz regime \cite{Li}.

\begin{figure}[h]
\centering
\includegraphics[width=8cm]{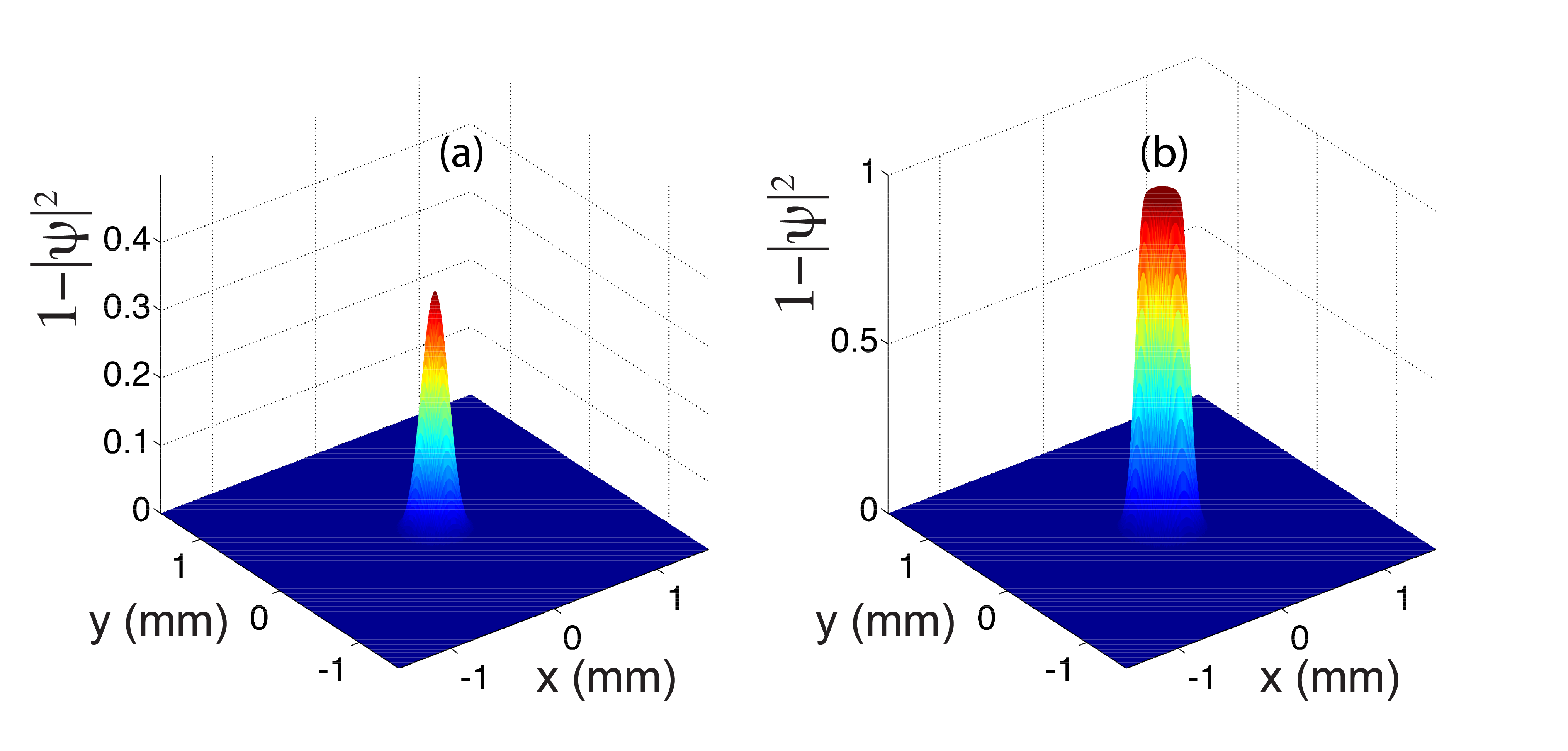}
\caption{(a) Early frame from the 2D simulation, taken at $t=-7.5$ ps [same horizontal scale used in Fig. \ref{fig2}(a)], showing the effect of a short sech-type pulse on the spatial distribution of order parameter; (b) same as (a) but for a later time $t=-4.5$ ps, showing the saturation of the order parameter around the highest intensity region of the beam, where the pulse has locally destroyed all the available Cooper pairs.}
\label{fig3}
\end{figure}

In the first simulation we show what happens in a simplified scenario, when taking into account one spatial dimension only ($x$). Figure \ref{fig2} displays the results of simulations in the 1D case analysing the effects of a short pulse of the form $A(x,t)\equiv A_{0}\mathrm{sech}\left( t/t_{0} \right) \mathrm{sin}\left( \omega_{0}t \right) e^{-(x/x_{0})^2}$ at intensity $I=16I_{0}=49.6$ kW/cm$^{2}$, and a beam width $x_{0}=200$ $\mu$m; as well as an 8-th order supergaussian pulse with the same duration and peak intensity. Fig. \ref{fig2}(a) shows the sech-type pulse intensity (normalised to unity, blue line) and its effect on the order parameter (quantity $1-|\psi|^2$, also normalized to unity, red line). Note that we show $1-|\psi|^{2}$ for clarity, so when this quantity increases the pulse is {\em destroying} Cooper pairs. It can be seen that the order parameter drops in step-like fashion with each peak and trough of the pulse, eventually recovering after interacting with the pulse (not shown). Fig. \ref{fig2}(b) presents the same data in this case for the supergaussian pulse, also showing a very similar dynamics. The supercurrents generated by both types of pulse are given in Fig. \ref{fig2}(c), whilst Fig. \ref{fig2}(d) displays the emission spectra induced by both pulses, showing the formation of odd harmonics in both cases, as expected from our theory.
We now consider two spatial dimensions ($x,y$) in Fig. \ref{fig3}, which displays the results of simulations in two spatial dimensions analysing the effects of a short pulse of the form $A(x,y,t)=A_{0}\mathrm{sech}\left( t/t_{0} \right) \mathrm{sin}\left( \omega_{0}t \right) e^{{-\left( x^2 +y^2\right)/x_{0}^{2}}}$, using the same parameters as in Fig. \ref{fig2}. Figures \ref{fig3}(a) and \ref{fig3}(b) show frames of the evolution of the order parameters at $t=-7.5$ ps and $t=-4.5$ ps using the same horizontal scale as in Fig. \ref{fig2}(a): Fig. \ref{fig3}(a) showing the dip in order parameter as the pulse first interacts with the material, while Fig. \ref{fig3}(b) showing the plateau as superconductivity is completely destroyed in a localised region about which pulse intensity is highest. Results in the 2D case exactly match those of the 1D case when the spatial distribution of the pulse is approximated at its origin $x=y=0$ value, and give the same spectra and supercurrents as in Fig. \ref{fig2}, confirming the validity and the accuracy of the simplified 1D model above.

\paragraph{Conclusions ---} 
It is found that type-II superconducting thin films below the critical temperature show an exceptionally large Kerr effect at low light intensities with a fast response on ultrafast time scales in the THz regime. Our simulations show in real-time the effects on the order parameter of Nb due to sech-type and supergaussian incident pulses and the efficient generation of odd harmonics. The approach used here is simple and effective, and is generally suited to the study of all type-II superconductors. We hope this work will stimulate further investigations into superconductor-based nonlinear optical materials for ultrafast applications, with these materials proving to be a key element in future nonlinear photonics systems. 

K.A.F. is supported by the Scottish Carnegie Trust.

\end{document}